# Heat Transfer in the Flow of a Cold, Axisymmetric Vertical Liquid Jet against a Hot, Horizontal Plate

**Jian-Jun SHU**[*] and **Graham WILKS**

School of Mechanical & Aerospace Engineering, Nanyang Technological University,

50 Nanyang Avenue, Singapore 639798

**ABSTRACT**: The paper considers the heat transfer characteristics of a thin film flow over a hot horizontal flat plate resulting from a cold vertical liquid jet falling onto a surface. A numerical solution of high accuracy is obtained for large Reynolds numbers using the modified Keller box method. For the flat plate, solutions for axisymmetric jets are obtained. In parallel approximation theory, an advanced polynomial approximation of the velocity and temperature distributions is employed, and results are in good agreement with those obtained using a simple Pohlhausen polynomial and numerical solutions.

**KEYWORDS**: Thin film flow, large Reynolds numbers and modified Keller box method.

## 1 Introduction

As previously pointed out, in technological processes involving heat or mass transfer, the draining flow of liquid through banks of horizontal tubes under the force of gravity often occurs. The mode of drainage may be in the form of droplets, columns or continuous sheets. After having examined the sheet mode of drainage it is natural to move on to a closer inspection of the columnar mode of drainage. Again, if the film thickness is small relative to a typical tube dimension, the impact surface may be regarded as locally plane [1]. Accordingly, an initial prototype model for columnar impingement is simply that of a vertical round jet striking a plane horizontal surface. Therefore, the flow model is axisymmetric and considerable simplification can be made to the governing equations. Some detailed understanding of the flow and heat transfer characteristics at the point of impingement may be obtained and methodologies identified for examining the non-axisymmetric flow at the right time.

---

[*] Author to whom correspondence should be addressed.

Analytically Watson [2] found a similarity solution of the boundary-layer equations governing such a flow and also considered by approximate methods the initial growth of the boundary layer from the stagnation point where the similarity solution does not hold. Chaudhury [3] obtained a supplementary thermal solution using an orthogonal polynomial. Some recent progress [1,4-9] has been made in investigating various problems of a liquid jet impinging on a solid surface. In this paper the theoretical results are improved, and an accurate numerical solution is obtained for the heat transfer in the flow of a cold, axisymmetric vertical liquid jet against a hot horizontal plate.

## Nomenclature

| | | | |
|---|---|---|---|
| $C_p$ | specific heat | Greek symbols | |
| $(f,u,v,\phi,w)$ | dependent variables | $\beta$ | dimensionless free surface temperature |
| $f_0,\phi_0$ | initial profiles | $\Gamma$ | Gamma function |
| $H,h$ | film thicknesses | $\Delta$ | $\delta_T/\delta$ |
| $H_0$ | jet semi-thickness | $\delta,\delta_T$ | dimensionless boundary layer thicknesses of velocity and temperature variations |
| $k$ | thermal conductivity | $\eta_T$ | $\eta/\Delta$ |
| $l$ | leading edge shift constant | $\kappa$ | thermometric conductivity |
| $N_u$ | Nusselt number | $\mu,\nu$ | dynamic and kinematic viscosities |
| $P,p$ | pressure | $(\xi,\eta)$ | dimensionless coordinates |
| $P_r$ | Prandtl number | $\rho$ | density |
| $Q$ | $U_0\,H_0$ | $\tau$ | skin friction |
| $q$ | heat flux | $\psi$ | stream function |
| $(r,z),(x,Y)$ | Cartesian coordinates measured along and perpendicular to plate | | |
| $R_e$ | Reynolds number | | |
| $T$ | temperature | | |
| $T_0$ | jet temperature | Subscripts | |
| $T_w$ | plate temperature | $w$ | Watson |
| $U_0$ | jet velocity | | |
| $\overline{U}_s$ | free surface velocity | | |
| $\vec{V}=(U,V)$ | velocity and its components | Superscript | |
| $x_0,x_1$ | ends of Regions $1$ and $2$ | $\bar{}$ | dimensional analysis |

## 2 Modelling

The problem to be examined concerns the film cooling which occurs when a cold vertically draining column strikes a hot horizontal plate. Although a column of fluid draining under gravity is accelerated and thin at impact [10,11], it is reasonable to model the associated volume flow as a jet of uniform velocity $U_0$, uniform temperature $T_0$ and radius $H_0$ as is illustrated in Figure 1(a). The notation $Q = \pi H_0^2 U_0$ is introduced for the flow rate and a film Reynolds number may be defined as $R_e = \frac{U_0 H_0}{\nu}$ where $\nu$ is the kinematic viscosity of the fluid. Watson [2] delineated the underlying hydrodynamics of fluid flow. Exactly the same physical assessment of the flow field applies as was outlined in [4]. The sub-regions (i)-(v) are once again appropriate and the solution, to follow, similarly uses this understanding of the basic hydrodynamics.

**Figure 1(a)**: Vertical jetting of an axisymmetric flat plate and resultant thin film

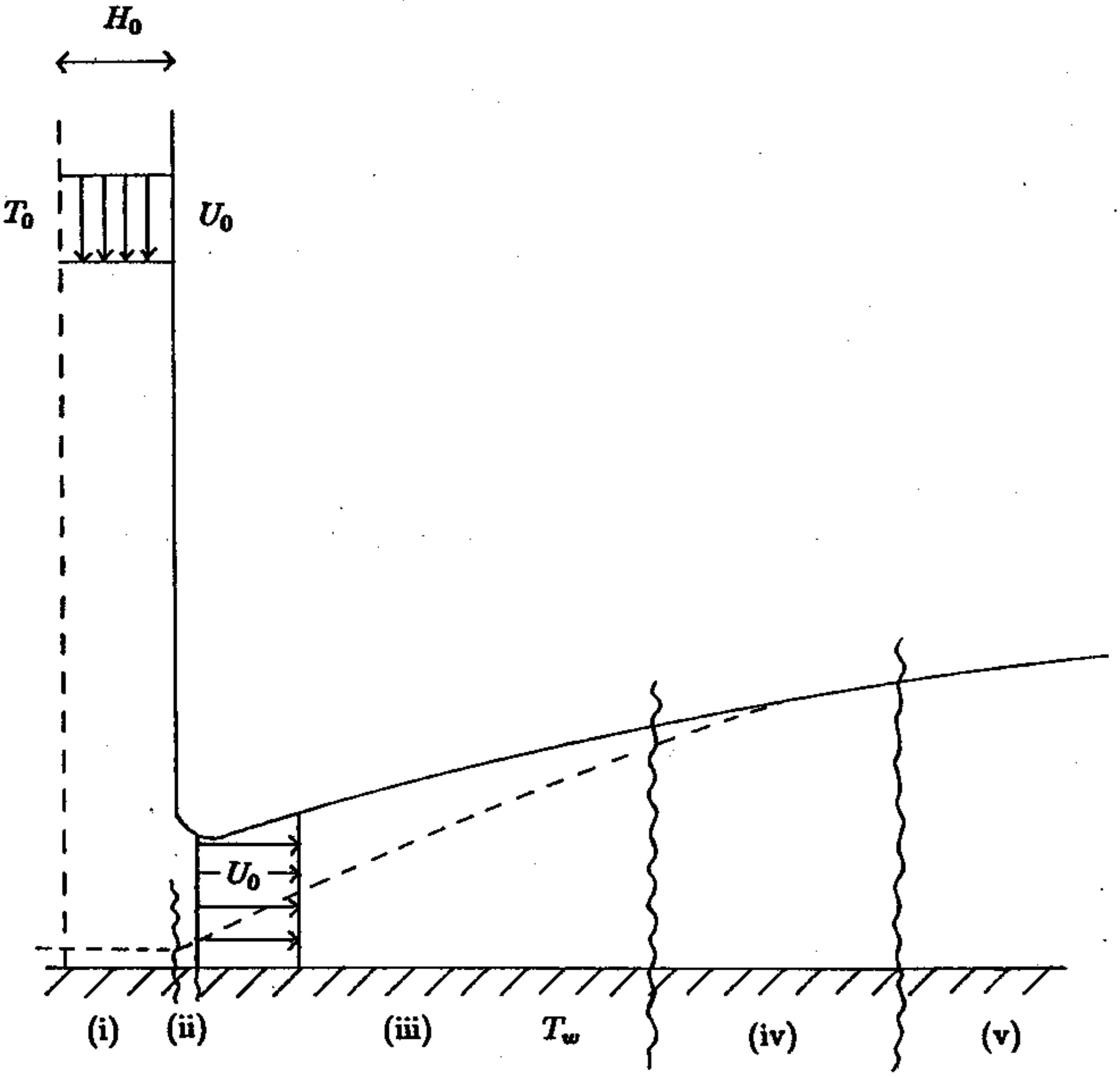

(i) Imbedded stagnation boundary layer
(ii) Outer inviscid deflection region
(iii) Quasi Blasius viscous diffusion
(iv) Transition around viscous penetration
(v) Similarity film flow

The dashed line represents the hydrodynamic boundary layer.

In practice, the flow against a plane wall is terminated by a hydraulic jump. The objective here is at the right time to develop a methodology for the flow around a tube where no such phenomenon is observed. Accordingly, the associated complications of a possible hydraulic jump will not be considered.

Heat transfer estimates can be obtained under the assumptions of a constant temperature $T_w$ at the plane and zero heat flux on the free surface. When water is the coolant medium then it has to be noted that the rates of viscous and thermal diffusion are appreciably different and the point at which viscous effects penetrate the free surface occur before the point at which the free surface first experiences the presence of the hot plane. To develop an approximation the flow field is thus divided into the following regions, as is illustrated in Figure 1(b):

**Figure 1(b)**: Basis of approximate solution

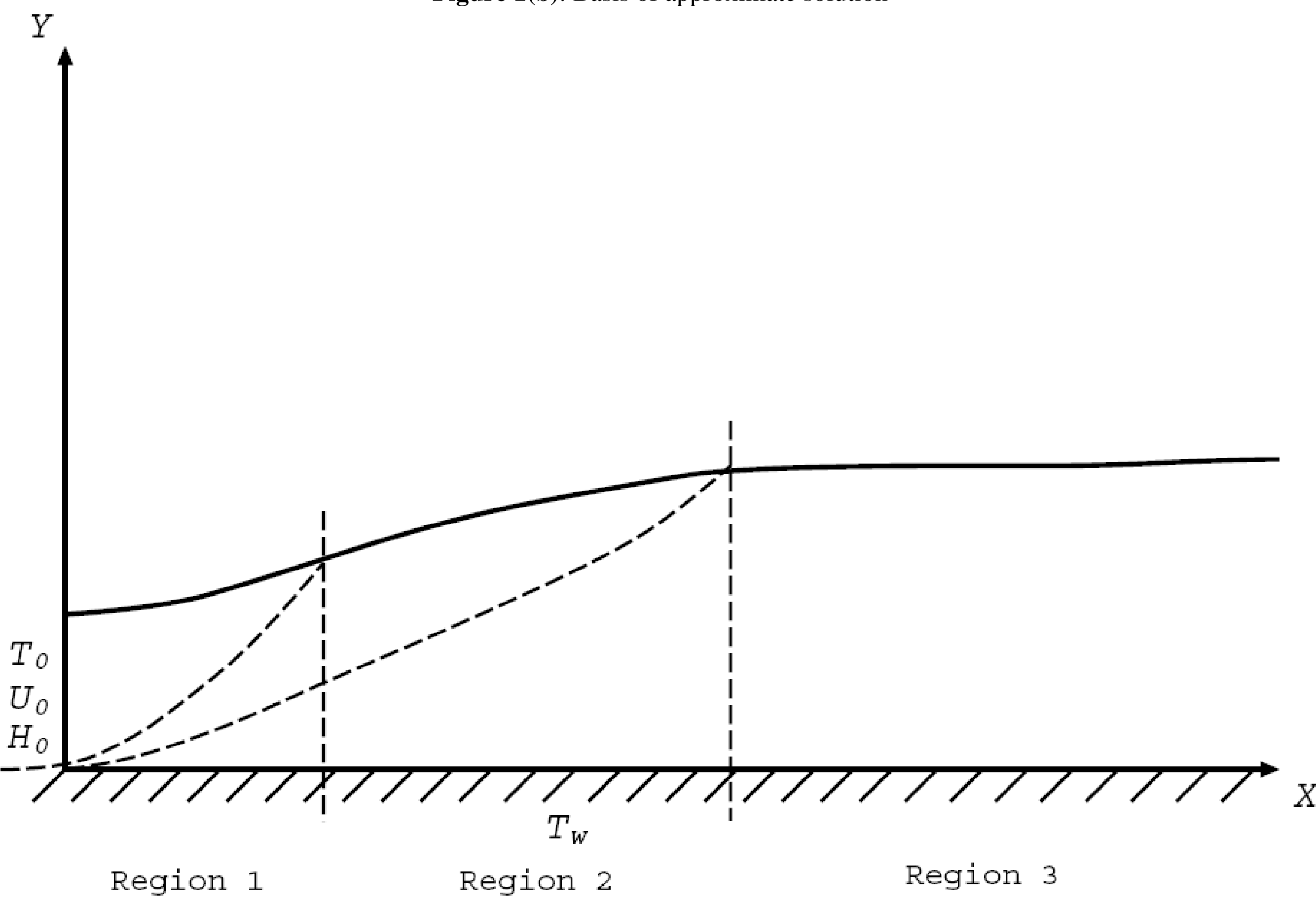

**Region 1**

When an impinging circular jet strikes a plane surface it experiences on eventually inviscid radially symmetric division and deflection through $90^o$. In the immediate vicinity of the point of impact viscous effects begin to influence the flow field. At that point, an imbedded axisymmetric stagnation boundary layer of thickness $O\left(\frac{\nu H_0}{U_0}\right)^{\frac{1}{2}}$ or $O\left(H_0 R_e^{-\frac{1}{2}}\right)$ occurs. For high flow rates $R_e >> 1$ and to a first approximation the stagnation boundary layer thickness is negligibly small. The presence of the solid boundary more significantly influences the flow within the deflected jet away from the point of impact. Here a viscous boundary layer develops against the horizontal plate. The effects of viscosity eventually are present throughout the film and the form of solution identifies a point of viscous penetration of the free surface marking the end of Region 1. For the Prandtl numbers greater than $1$ a parallel thermal diffusion occurs but thermal effects do not penetrate the free surface at the end of Region 1.

## Region 2

Over Region 2 thermal diffusion continues until the presence of the hot plate eventually influences the free surface. This point of penetration marks the end of Region 2. Accordingly, Region 2 describes the developing temperature distribution within an established hydrodynamic flow field since the form of solution in Region 1 is designed to coalesce directly with the Watson similarity solution for velocity and film thickness at the boundary between Region 1 and 2.

## Region 3

Once the influence of the hot plate penetrates the free surface the surrounding air acts as an insulator and here there is zero heat flux from the free surface. As a consequence, the film gradually approaches a uniform temperature distribution coinciding with the temperature of the plate. Region 3 monitors the approach to this asymptotic state.

To assess the validity of the approximate method a full numerical solution of the model equations is then obtained. Results and comparisons are presented at the conclusion of the paper.

## 3 Governing Equations

The flow under investigation has been modelled as a steady, axisymmetric flow of incompressible fluid. In the absence of body forces and viscous dissipation, the equations expressing the conservation of mass, momentum and energy are consequently

$$\nabla \bullet \vec{V} = 0 \tag{1}$$

$$\rho\left(\vec{V} \bullet \nabla\right)\vec{V} = -\nabla P + \mu \nabla^2 \vec{V} \tag{2}$$

$$\rho C_p \left(\vec{V} \bullet \nabla\right) T = k \nabla^2 T \tag{3}$$

where $\vec{V} = \left(v_r, v_z\right)$ are velocity components associated with cylindrical coordinates $\left(r, z\right)$ measured along the plate from the axis of deflection and perpendicular to the plate, respectively. $\rho$, $\mu$, $C_p$ and $k$ are the density, dynamic viscosity, specific heat at constant pressure and the thermal conductivity of the cooling fluid in the jet, respectively. $T$ and $P$ are, respectively, the temperature and pressure within the fluid.

The specific boundary conditions under which the equations are to be solved closely parallel those of [4]. In particular, at the wall, the no slip condition and the constant temperature $T_w$ require

$$v_r = v_z = 0,\ T = T_w \quad \text{on} \quad z = 0,\ r \geq 0. \tag{4}$$

On the free surface, assuming negligible shear stress and heat flux, requires

$$\frac{\partial v_r}{\partial z} + \frac{\partial v_z}{\partial r} = 0,\ \frac{\partial T}{\partial z} = 0 \quad \text{at} \quad z = H\left(r\right),\ r \geq 0. \tag{5}$$

The conservation of volume constraint applies at any given $r$ station and hence

$$2\pi \int_0^{H(r)} r\, v_r\left(r, z\right) dz = \text{constant} = \pi H_0^2 U_0 \quad \text{for} \quad r \geq 0. \tag{6}$$

Under the assumption that the film thickness remains thin relative to a characteristic horizontal dimension, a boundary layer treatment of the equations leads to significant simplification. The following non-dimensional variables are introduced

$$x = \frac{r}{R_e^{\frac{1}{3}} H_0}, \quad \overline{Y} = \frac{R_e^{\frac{1}{3}} z}{H_0}, \tag{7}$$

$$\overline{H}(x) = \frac{R_e^{\frac{1}{3}} H(r)}{H_0}, \tag{8}$$

$$\overline{U} = \frac{v_r}{U_0}, \quad \overline{V} = \frac{R_e^{\frac{2}{3}} v_z}{U_0}, \quad \overline{\phi} = \frac{T - T_w}{T_0 - T_w}, \quad p = \frac{P}{\rho U_0^2}. \tag{9}$$

In the limit $R_e \to +\infty$ with $x$ remaining $O(1)$, the following equations are obtained

$$\frac{\partial}{\partial x}\left(x\overline{U}\right) + \frac{\partial}{\partial \overline{Y}}\left(x\overline{V}\right) = 0 \tag{10}$$

$$\overline{U}\frac{\partial \overline{U}}{\partial x} + \overline{V}\frac{\partial \overline{U}}{\partial \overline{Y}} = -\frac{\partial p}{\partial x} + \frac{\partial^2 \overline{U}}{\partial \overline{Y}^2} \tag{11}$$

$$0 = \frac{\partial p}{\partial \overline{Y}} \tag{12}$$

$$P_r\left(\overline{U}\frac{\partial \overline{\phi}}{\partial x} + \overline{V}\frac{\partial \overline{\phi}}{\partial \overline{Y}}\right) = \frac{\partial^2 \overline{\phi}}{\partial \overline{Y}^2} \tag{13}$$

where $P_r = \frac{\nu}{\kappa}$ is the Prandtl number with $\nu$ the kinematic viscosity $\frac{\mu}{\rho}$ and $\kappa$ the thermometric conductivity $\frac{k}{\rho C_p}$. In common with standard boundary layer theory (12), implying that the pressure across the film remains constant. In the absence of external pressure gradients and with zero shear assumed on the free surface, the pressure term in (11) is identically zero. In non-dimensional variables the boundary conditions now read

$$\overline{U} = \overline{V} = \overline{\phi} = 0 \quad \text{on} \quad \overline{Y} = 0, \; x \geq 0 \tag{14}$$

$$\frac{\partial \overline{U}}{\partial \overline{Y}} = \frac{\partial \overline{\phi}}{\partial \overline{Y}} = 0 \quad \text{at} \quad \overline{Y} = \overline{H}(x), \; x \geq 0 \tag{15}$$

$$\int_0^{\overline{H}(x)} x\overline{U}\, d\overline{Y} = \frac{1}{2} \quad \text{for} \quad x \geq 0. \tag{16}$$

These have been quoted in the context of the fully developed film flow field, which is approached in Region 3. Watson [2] showed that the hydrodynamic equations for this system possess similarity solutions. A simple supplementary thermal solution of uniform temperature is also present. These solutions provide the basis for developing comprehensive approximations for the complete flow field downstream of the radial symmetry point of impingement incorporating Regions 1, 2 and 3.

## 4 Downstream Similarity Solutions

The equations to be solved are

$$\frac{\partial}{\partial x}\left(x\overline{U}\right)+\frac{\partial}{\partial \overline{Y}}\left(x\overline{V}\right)=0 \tag{17}$$

$$\overline{U}\,\frac{\partial \overline{U}}{\partial x}+\overline{V}\,\frac{\partial \overline{U}}{\partial \overline{Y}}=\frac{\partial^2 \overline{U}}{\partial \overline{Y}^2} \tag{18}$$

$$P_r\left(\overline{U}\,\frac{\partial \overline{\phi}}{\partial x}+\overline{V}\,\frac{\partial \overline{\phi}}{\partial \overline{Y}}\right)=\frac{\partial^2 \overline{\phi}}{\partial \overline{Y}^2} \tag{19}$$

subject to boundary conditions (14)-(16). Note that the equations governing the hydrodynamics (17)-(18) are independent of the energy equation (19). These are examined as follows. It is well known that, using the Mangler's transformations [12], in an infinitely expansive fluid, the calculation of an axisymmetric boundary layer on a rotating body can be reduced to that of a complementary two-dimensional flow. The flow currently under consideration represents the axially symmetric equivalent of the two-dimensional flow examined in [4]. Here we examine the possibility that the Mangler's transformations applied to the set of equations (17)-(18) under the specified hydrodynamic boundary conditions have a two-dimensional equivalent differential system. Let

$$x'=\int_0^x t^2\,dt=\frac{x^3}{3} \tag{20}$$

and

$$\overline{Y}'=x\overline{Y} \tag{21}$$

then

$$\overline{U}' = \overline{U} \tag{22}$$

$$\overline{V}' = \frac{1}{x}\left(\overline{V} + \frac{\overline{Y}\overline{U}}{x}\right). \tag{23}$$

With these new $'$ variables the equations (17)-(18) become

$$\frac{\partial \overline{U}'}{\partial x'} + \frac{\partial \overline{V}'}{\partial \overline{Y}'} = 0 \tag{24}$$

$$\overline{U}'\frac{\partial \overline{U}'}{\partial x'} + \overline{V}'\frac{\partial \overline{U}'}{\partial \overline{Y}'} = \frac{\partial^2 \overline{U}'}{\partial \overline{Y}'^2}. \tag{25}$$

The boundary conditions are

$$\overline{U}' = \overline{V}' = 0 \quad \text{on} \quad \overline{Y}' = 0 \tag{26}$$

$$\frac{\partial \overline{U}'}{\partial \overline{Y}'} = 0 \quad \text{on} \quad \overline{Y}' = x\overline{H}(x) = \overline{H}'(x') \quad \text{say} \tag{27}$$

and

$$\int_0^{\overline{H}'(x')} \overline{U}' \, d\overline{Y}' = \frac{1}{2}. \tag{28}$$

With the minor modification of the conservation of volume flow constraint the differential system formally reproduces the hydrodynamic system examined in [4]. Introducing a similarity variable $\eta = \overline{Y}'/\overline{H}'(x')$, a stream function form of solution $\psi(x,\overline{Y}') = \overline{U}_s(x)\overline{h}(x)f(\eta)$ leads to the Watson similarity solution as the solution of

$$2f''' + 3c^2 f'^2 = 0 \qquad f(0) = 0 \quad f(1) = 1 \quad f''(1) = 0.$$

Here $\overline{U}_s(x)$ represents the non-dimensional unknown velocity at the free surface and $c$ can be obtained analytically as

$$\frac{\Gamma\left(\frac{1}{3}\right)\Gamma\left(\frac{1}{2}\right)}{3\Gamma\left(\frac{5}{6}\right)} \approx 1.402.$$

As a result, the axisymmetric solution can be directly inferred from the solution of [4] as

$$\overline{U}_s(x) = \frac{27c^2}{8\pi^2\left(x^3 + l^3\right)} \tag{29}$$

$$\overline{h}(x) = \frac{2\pi}{3\sqrt{3}}\left(x^3 + l^3\right). \tag{30}$$

Here $l$ is a non-dimensional leading edge shift constant. The solutions hold at large distances from the jet incidence and $l$ may be associated with an indeterminate origin of such a flow solution. An estimate of $l$ can be obtained by further considering the growth of the boundary layer from the point of jet impact. The effects of viscous retardation are seen to result in a simultaneous thickening of the film and a reduction in the free surface velocity. In the original dimensional variables

$$U_s(r) = \frac{27c^2}{8\pi^4}\frac{Q^2}{\nu\left(r^3 + L^3\right)} \tag{31}$$

$$H(r) = \frac{2\pi^2}{3\sqrt{3}}\frac{\nu}{Q}\frac{\left(r^3 + L^3\right)}{r}. \tag{32}$$

The velocity distribution within the film is given by

$$v_r(r,z) = U_s(r)f'(\eta) \tag{33}$$

where $f(\eta)$ is the Watson similarity solution whose properties have been presented in [4]. The associated asymptotic downstream similarity solution for the temperature distribution is obtained by examining equation (19) with $\overline{\phi}\left(x,\overline{Y}\right) = \overline{\phi}(\eta)$ together with the associated similarity transformations. The resultant equation is

$$\overline{\phi}'' = 0 \tag{34}$$

subject to the boundary conditions

$$\overline{\phi}(0) = 0,\ \overline{\phi}'(1) = 0\,. \tag{35}$$

The solution must therefore be $\overline{\phi} \equiv 0$ confirming a uniform temperature distribution in the film equal to the temperature of the plate, $T_w$.

# 5 Approximate Solutions

The solutions obtained in the previous section are asymptotic and they are valid well downstream of the location of jet impingement and deflection along the plane. An approximation scheme is now presented which looks more closely at the flow at impingement. The solution is built up from this vicinity, stage by stage, to provide comprehensive details of the velocity and temperature distribution at radial stations away from the origin.

## 5.1 Region 1

The discussion of the Region follows closely that of the sheet drainage flow of [4]. At impact, an inviscid deflection of the draining sheet occurs over a negligibly small length scale. Only after deflection is the flow aware of the presence of the solid boundary and only then do viscous effects begin to influence the flow field. The development of a viscous boundary layer within a uniform velocity film accounts for the close parallel in this region with Blasius boundary layer flow. Similarly, the temperature differential between the plane and the fluid only begins to influence the temperature distribution after deflection. Thus, a developing thermal boundary layer may also be anticipated from $r = 0$. The equations governing the viscous and thermal boundary layers are exactly the same as (17)-(19) but the boundary conditions, as in [4], now read

$$\overline{U} = 0,\ \overline{V} = 0,\ \overline{\phi} = 0 \quad \text{on} \quad \overline{Y} = 0,\ x \geq 0$$

$\overline{U} \to 1,\ \overline{\phi} \to 1$ as $\overline{Y}$ approaches the outer limits of the viscous and thermal boundary layers, respectively

$$\overline{U} = 1,\ \overline{\phi} = 1 \quad \text{at} \quad x = 0,\ \overline{Y} > 0.$$

The transformations

$$\psi\left(x,\overline{Y}\right) = \sqrt{\frac{2}{3}}\,x^{\frac{3}{2}}\,\overline{f}\left(\overline{\eta}\right),\ \overline{\phi}\left(x,\overline{Y}\right) = \overline{\phi}\left(\overline{\eta}\right),\ \overline{\eta} = \sqrt{\frac{3}{2}}\,\frac{\overline{Y}}{x^{\frac{1}{2}}}$$

$$\overline{U} = \frac{1}{x}\frac{\partial\psi}{\partial\overline{Y}},\ \overline{V} = -\frac{1}{x}\frac{\partial\psi}{\partial x} \tag{36}$$

lead once again to

$$\overline{f}''' + \overline{f}\overline{f}'' = 0$$

$$\frac{1}{P_r}\bar{\phi}''+\bar{f}\bar{\phi}'=0$$

subject to boundary conditions

$$\bar{f}(0)=0,\ \bar{f}'(0)=0,\ \bar{\phi}(0)=0$$

$$\bar{f}'(\bar{\eta})\to 1,\ \bar{\phi}(\bar{\eta})\to 1 \quad \text{as } \bar{\eta}\to +\infty .$$

Full details of these solutions can be found in [4]. Following the arguments of [4] a device which suppresses the transition region may be introduced. An approximate velocity profile

$$\overline{U}(x,\overline{Y})=\overline{U}_s(x)f'\left(\frac{\overline{Y}}{\delta}\right),\ \eta=\frac{\overline{Y}}{\delta(x)} \tag{37}$$

is assumed, where $f'(\eta)$ is the original Watson similarity profile and $\delta(x)$ is the non-dimensional boundary layer thickness. The profile is then used in a Kármán-Pohlhausen method of solution. Over Region 1 unretarded fluid is present when $x<x_0$, say where $x_0$ marks the point of penetration of viscous effects on the free surface, so that $\overline{U}_s(x)=1$ and $\delta(x)<\bar{h}(x)$ over $0<x<x_0$. For $x>x_0$ into Region 2 $\delta(x)\equiv\bar{h}(x)$ and $\overline{U}_s(x)<1$ in a manner which, using the conservation of flow constraint, can be matched directly onto the asymptotic similarity solutions. The use of $f'(\eta)$ is feasible only because its integral properties are readily available for use in the momentum integral equation, which reads

$$\left(\frac{1}{x^2}\right)\frac{d}{dx}\int_0^{\delta(x)}\overline{U}\left(1-\overline{U}\right)d\overline{Y}=\left(\frac{\partial\overline{U}}{\partial\overline{Y}}\right)_{\overline{Y}=0}. \tag{38}$$

Using (37) gives

$$\left(\frac{1}{x^2}\right)\frac{d}{dx}\left\{\delta(x)\int_0^1 f'(1-f')d\eta\right\}=\frac{f''(0)}{\delta(x)}=\frac{c}{\delta(x)}.$$

The integral

$$\int_0^1\left(f'-f'^2\right)d\eta=\frac{2\left(\pi-c\sqrt{3}\right)}{3\sqrt{3}c^2} \tag{39}$$

and the equation for the boundary layer thickness is

$$\frac{d}{dx}\left(\delta^2\right)=\frac{3\sqrt{3}c^3x^2}{\pi-c\sqrt{3}}$$

and hence

$$\delta^2(x)=\frac{\sqrt{3}c^3x^3}{\pi-c\sqrt{3}} \tag{40}$$

where $\delta(x)=0$ has been assumed at $x=0$ which is valid in the limit of the underlying assumption. Invoking the conservation of volume flow at $x_0$, the end point of Region 1, effectively suppresses the transition region and leads to

$$\int_0^{\delta(x)}\overline{U}\,d\overline{Y}+\left(\overline{h}-\delta\right)=\frac{1}{2} \tag{41}$$

or

$$\delta\int_0^1 f'(\eta)d\eta+\left(\overline{h}-\delta\right)=\frac{1}{2} \tag{42}$$

giving

$$\overline{h}(x)=\frac{1}{2}+\left(1-\frac{2\pi}{3\sqrt{3}c^2}\right)\delta\,. \tag{43}$$

Since $\delta(x_0)=\overline{h}$

$$x_0=\sqrt[3]{\frac{9\sqrt{3}c\left(\pi-c\sqrt{3}\right)}{16\pi^2}}\approx 0.46216 \tag{44}$$

and matching the free surface velocity at $x=x_0$ leads to

$$l=\sqrt[3]{\frac{9\sqrt{3}c}{16\pi^2}\left(3\sqrt{3}c-\pi\right)}\approx 0.8308\,. \tag{45}$$

## 5.2 Alternative Profiles

The fact that the integral properties of the Watson similarity profile are readily calculated has been used to advantage in 5.1. Its use however in the energy integral equation is not as convenient as a polynomial representation or approximation for the velocity profile. For example, the Pohlhausen profile

$$f_p'(\eta) = 2\eta - 2\eta^3 + \eta^4 \tag{46}$$

can be used again to evaluate aggregate properties of the flow. The result is an estimate of the boundary layer thickness given by

$$\delta_p^2 = \frac{420x^3}{37} \tag{47}$$

which leads to

$$x_{0\,p} = \frac{\sqrt[3]{3330}}{42} \approx 0.3555 \tag{48}$$

$$l_p = \sqrt[3]{\frac{27783c^2 - 370\pi^2}{8232\pi^2}} \approx 0.8560 \tag{49}$$

as compared to (44)-(45). Alternatively, the fourth-order polynomial approximation to $f'(\eta)$

$$f_w'(\eta) = c\eta + (4-3c)\eta^3 + (2c-3)\eta^4 \tag{50}$$

may be used. The viscous boundary layer thickness for this profile is given by

$$\delta_w^2 = \frac{420cx^3}{72 + 39c - 19c^2} \tag{51}$$

and

$$x_{0\,w} = \sqrt[3]{\frac{5(72 + 39c - 19c^2)}{21c(8+3c)^2}} \approx 0.46697 \tag{52}$$

$$l_w = \sqrt[3]{\frac{567c^3(8+3c)^2 - 40\pi^2(72 + 39c - 19c^2)}{168\pi^2 c(8+3c)^2}} \approx 0.8293 \tag{53}$$

which very closely approximate (44)-(45). Consequently, this polynomial $f_w'(\eta)$ is used for the subsequent development of velocity and temperature distributions. The temperature characteristics of Region 1 are now considered. The energy integral equation of (19) becomes

$$\left(\frac{1}{x^2}\right)\frac{d}{dx}\int_0^{\delta_T(x)} \overline{U}\left(1-\overline{\phi}\right)d\overline{Y} = \frac{1}{P_r}\left(\frac{\partial\overline{\phi}}{\partial\overline{Y}}\right)_{\overline{Y}=0} \tag{54}$$

where $\delta_T(x)$ denotes the outer limits of the region of thermal diffusion. When $P_r > 1$, it may be anticipated that $\delta_T(x) < \delta(x)$ over $0 < x < x_0$. With $\eta_T = \dfrac{\overline{Y}}{\delta_T(x)}$ and the ratio $\dfrac{\delta_T}{\delta}$ denoted by $\Delta$ so that $\eta = \Delta\eta_T$ the solution for $\delta_T(x)$ may be developed by assuming profiles for $\overline{U}$ and $\overline{\phi}$. The two profile pairs of [4] have been used

$$\text{(a)} \quad \overline{U}(\eta) = f_p'(\eta) \qquad \text{(b)} \quad \overline{U}(\eta) = f_w'(\eta)$$

$$\overline{\phi}(\eta_T) = f_p'(\eta_T) \qquad \overline{\phi}(\eta_T) = f_w'(\eta_T) \tag{55}$$

as each pairing ensures identical velocity and temperature distributions for $P_r = 1$ when also $\Delta = 1$. It is expected that the second pairing has advantages in effecting the transition at the end of Region 1. The resultant equations for $\Delta$ are

$$\text{(a)} \quad \Delta^3\left(168 - 27\Delta^2 + 7\Delta^3\right) = \frac{148}{P_r} \tag{56}$$

$$\text{(b)} \quad \Delta^2 D(\Delta) = \frac{4\left(72 + 39c - 19c^2\right)}{P_r} \tag{57}$$

where $D(\Delta) = 168c(3-c)\Delta + 27(4-3c)(5-2c)\Delta^3 - 7(3-2c)(12-5c)\Delta^4$ as obtained previously.

## 5.3 Region 2

The continuing diffusion of hot wall effects within a hydrodynamic setting prescribed by the Watson similarity solution is monitored in Region 2. The velocity on the free surface is no longer uniform but is prescribed in non-dimensional terms by (29). The viscous boundary layer thickness $\delta(x)$ is now the same as $\overline{h}(x)$, namely,

$$\delta(x) = \overline{h}(x) = \frac{2\pi}{3\sqrt{3}}\left(x^3 + l_w^3\right).$$

The energy integral equation (54) remains appropriate, and the progressive thermal diffusion implies that $\delta_T(x) \to \delta(x) = \overline{h}(x)$. In prescribing profiles $\eta_T = \dfrac{\overline{Y}}{\delta_T(x)}$ may again be utilised but now $\Delta(x) = \dfrac{\delta_T(x)}{\delta(x)}$ is no longer constant and tends to $1$ at the end of Region 2. An equation for $\delta_T(x)$ may be obtained by introducing the following profiles into the energy equation

$$\overline{U}(x,\eta) = \overline{U}_s(x) f_w'(\eta)$$

$$\overline{\phi}(x,\eta_T) = f_w'(\eta_T), \tag{58}$$

to give

$$\frac{\delta_T(x)}{x^2}\frac{d}{dx}\left[\overline{U}_s(x) D(\Delta)\delta_T(x)\right] = \frac{2520c}{P_r} \tag{59}$$

which in turn leads to

$$\frac{d}{dx}\left(\Delta^2\right) = \frac{10080x^2}{cP_r\left(x^3 + l_w^3\right)\Delta\left[336c(3-c) + 108(4-3c)(5-2c)\Delta^2 - 35(3-2c)(12-5c)\Delta^3\right]}. \tag{60}$$

This first-order equation may now be integrated with initial data $\Delta(x_{0w}; P_r)$ as far as $\Delta(x_{1w}(P_r); P_r) = 1$.

$$\begin{aligned}&\Delta^3\left[3360c(3-c) + 648(4-3c)(5-2c)\Delta^2 - 175(3-2c)(12-5c)\Delta^3\right]\\ &= \frac{50400}{cP_r}\ln\frac{x^3 + l_w^3}{x_{1w}^3 + l_w^3} + 6660 + 2001c - 1222c^2 .\end{aligned} \tag{61}$$

where $x_{1w}$ is used to denote the end of Region 2 as predicted using the Watson polynomial profile. Beyond $x_{1w}$ viscous and thermal effects are present throughout the film. For comparison, the Pohlhausen equivalent has also been computed and included in the subsequent illustration of results. The values of $x_{1w}(P_r)$ are listed in Table 1. The numerical details for various $P_r$ are presented in Table 2.

**Table 1**: Values of $x_{1w}(P_r)$ for various Prandtl numbers

| $P_r$ | $x_{1w}(P_r)$ |
|---|---|
| 1.0 | 0.4670 |
| 2.0 | 0.6274 |
| 3.0 | 0.7513 |
| 4.0 | 0.8616 |
| 5.0 | 0.9663 |
| 6.0 | 1.0691 |
| 7.0 | 1.1722 |
| 8.0 | 1.2773 |
| 9.0 | 1.3856 |
| 10.0 | 1.4980 |

**Table 2**: Numerical results for $\Delta(x)$ in Region 2 for various Prandtl numbers

| $\dfrac{x - x_{0w}}{x_{1w} - x_{0w}}$ | $\Delta(x)$ $P_r = 2$ | $P_r = 5$ | $P_r = 10$ |
|---|---|---|---|
| 0.0 | 0.7878 | 0.5775 | 0.4575 |
| 0.1 | 0.8092 | 0.6253 | 0.5347 |
| 0.2 | 0.8306 | 0.6728 | 0.6085 |
| 0.3 | 0.8520 | 0.7193 | 0.6769 |
| 0.4 | 0.8734 | 0.7644 | 0.7389 |
| 0.5 | 0.8947 | 0.8081 | 0.7948 |
| 0.6 | 0.9160 | 0.8500 | 0.8450 |
| 0.7 | 0.9372 | 0.8902 | 0.8899 |
| 0.8 | 0.9583 | 0.9286 | 0.9304 |
| 0.9 | 0.9792 | 0.9652 | 0.9669 |
| 1.0 | 1.0000 | 1.0000 | 1.0000 |

### 5.4 Region 3

In both Region 1 and Region 2 it has been assumed that at the edge of the developing thermal boundary layer the temperature smoothly assimilates into that of the impinging jet. As a consequence, zero heat flux has been invoked. In Region 3 the same boundary condition in fact remains valid. Here however it is justified by the assumption of negligible heat transfer between the liquid free surface and the surrounding air. Consequently, the temperature of the film rises as a result of continuing heat input at the plate. The temperature is thus asymptotic to $T_w$. Using the following profiles

$$\overline{U}(x,\eta) = \overline{U}_s(x) f_w'(\eta)$$

$$\overline{\phi}(x,\eta) = \beta(x) f_w'(\eta) \tag{62}$$

where now $\eta = \dfrac{\overline{Y}}{\overline{h}(x)}$, the energy integral equation now reads

$$\left(\frac{1}{x^2}\right)\left[\frac{d}{dx}\int_0^{\overline{h}(x)}\overline{U}\left(\beta-\overline{\phi}\right)d\overline{Y}-\int_0^{\overline{h}(x)}\overline{U}\,\frac{d\beta}{dx}\,d\overline{Y}\right]=\frac{1}{P_r}\left(\frac{\partial\overline{\phi}}{\partial\overline{Y}}\right)_{\overline{Y}=0}. \tag{63}$$

Here $\beta(x)$ monitors the adjustment of the film temperature to $T_w$. The result is an equation for $\beta(x)$ within the framework of prescribed film thickness, namely,

$$\frac{72+39c-19c^2}{630}\frac{d}{dx}\left[\overline{U}_s\overline{h}\overline{\beta}\right]-\frac{8+3c}{20}\overline{U}_s\overline{h}\frac{d\overline{\beta}}{dx}=\frac{c}{P_r}\frac{\beta x^2}{\overline{h}}. \tag{64}$$

Remembering that $\overline{U}_s(x)\overline{h}(x)=\dfrac{3\sqrt{3}c^2}{4\pi}$ equation (64) becomes

$$\frac{d\beta}{dx}=-\frac{2520\beta x^2}{c\left(360+111c+38c^2\right)P_r\left(x^3+l_w^3\right)} \tag{65}$$

and hence

$$\beta(x)=\left(\frac{x_{1\,w}^3+l_w^3}{x^3+l_w^3}\right)^{\frac{840}{c\left(360+11c+38c^2\right)P_r}}\approx\left(\frac{x_{1\,w}^3+l_w^3}{x^3+l_w^3}\right)^{\frac{1.015}{P_r}} \tag{66}$$

which satisfies the requirements $\beta\left(x_{1w}\left(P_r\right)\right)=1$ and has $\beta\to 0$ at rates dependent on $P_r$.

# 6 Approximation Results

The outlined approximation scheme provides comprehensive details of the flow and heat transfer characteristics for the modeled flow. Estimates of film thickness, velocity and temperature distributions, skin friction and heat transfer coefficients can be obtained for the entire region downstream of the point of impingement.

## 6.1 Film thickness, velocity and temperature distributions

To indicate the underlying implications of the hydrodynamic modelling the free surface velocity has been illustrated for the respective profiles in Figure 2(a). The associated velocity profile development within the deflected film is illustrated schematically in Figure 3. The overall prediction of film thicknesses is shown in Figure 4(a). A more detailed indication of the region by region form of solution appears in Figure 5. For the range of the Prandtl numbers $P_r = 2, 5$ and $10$ film thickness profiles incorporating the viscous and thermal diffusion processes to penetration are presented. A typical set of temperature distributions within the film covering the evolution between initial thermal diffusion and the asymptotic linear profile is illustrated in Figure 6.

**Figure 2(a)**: Comparison of free surface velocity for respective profiles

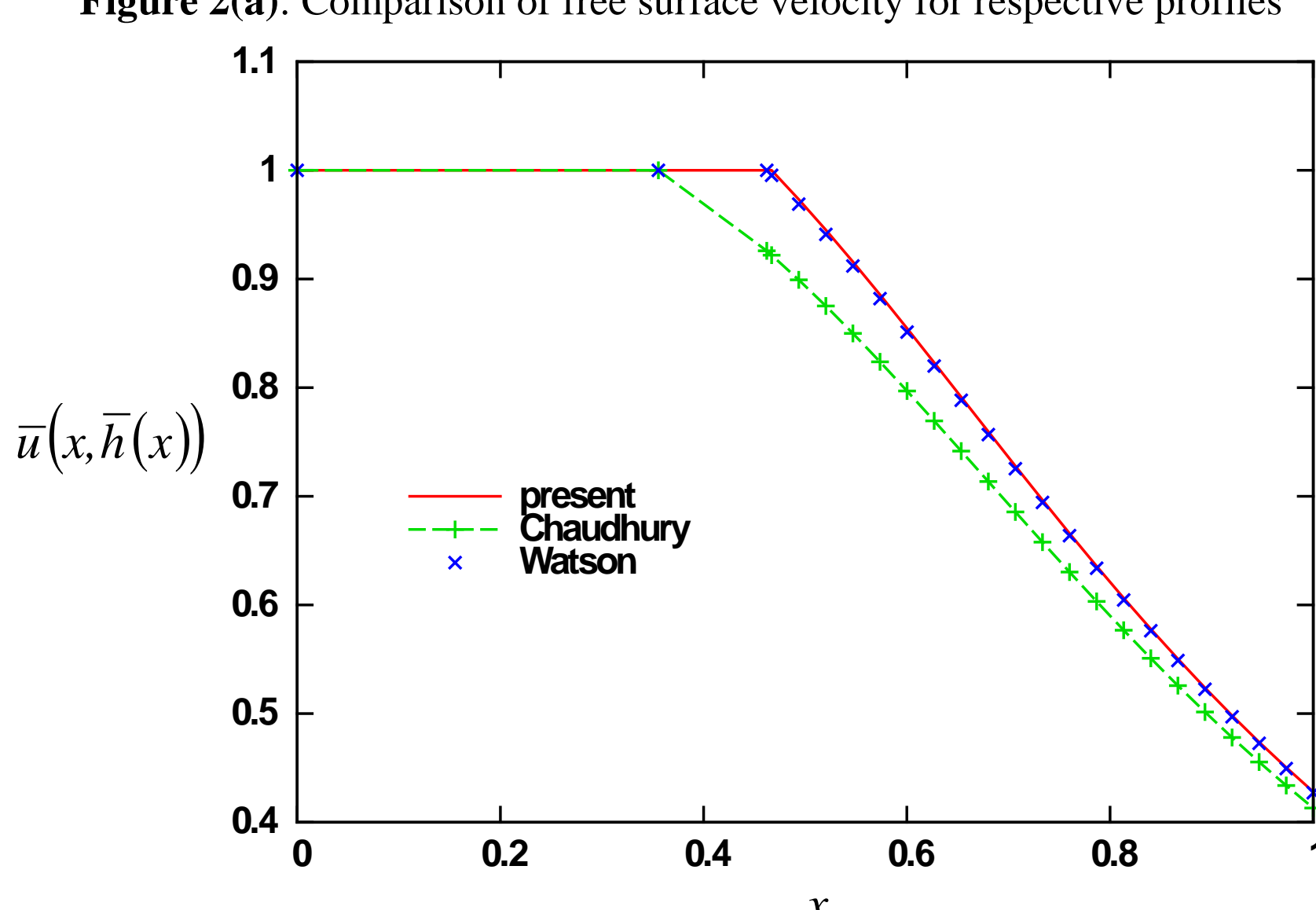

**Figure 2(b)**: Free surface velocity for numerical solution and present profile

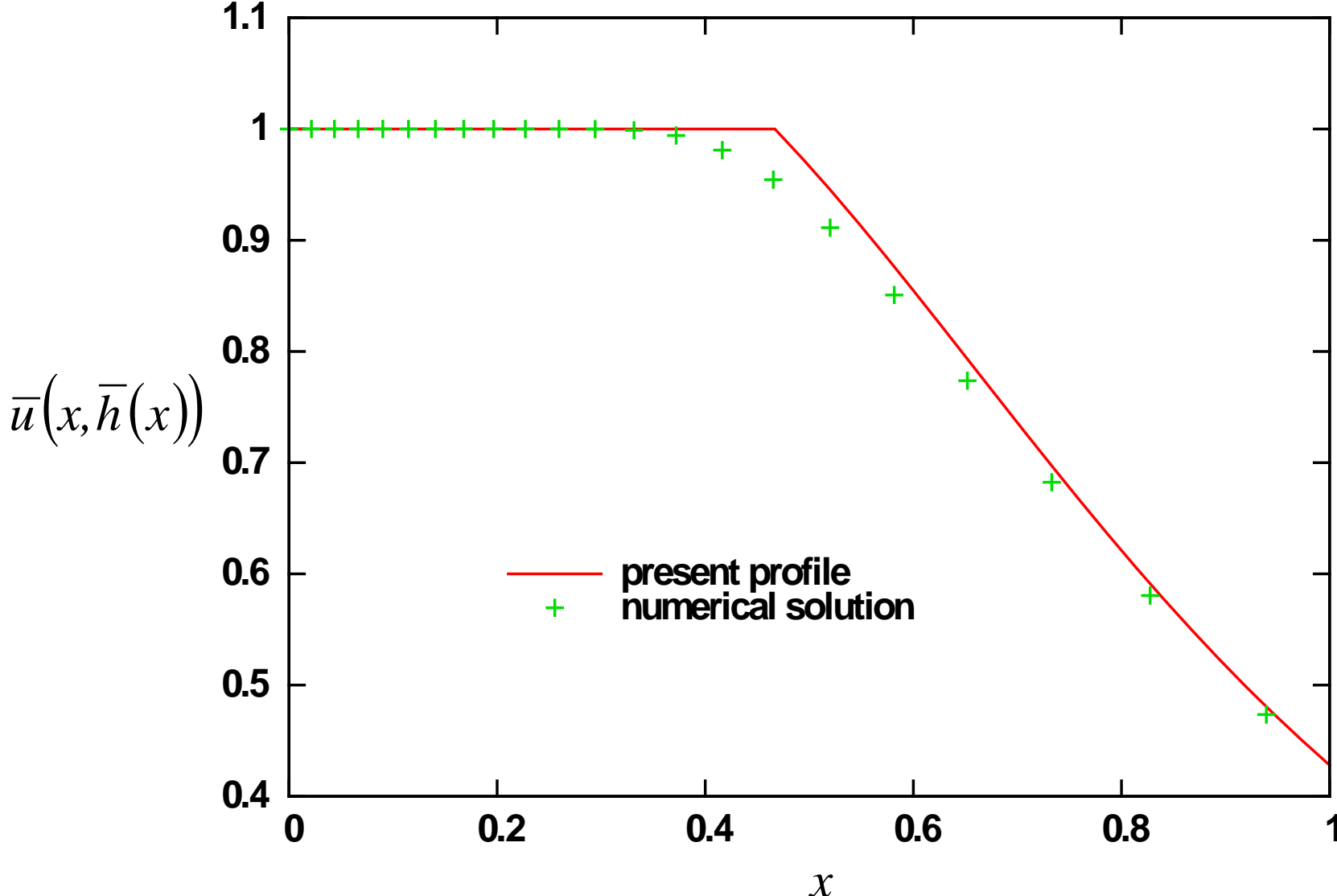

**Figure 3**: Velocity profile development within a deflection film

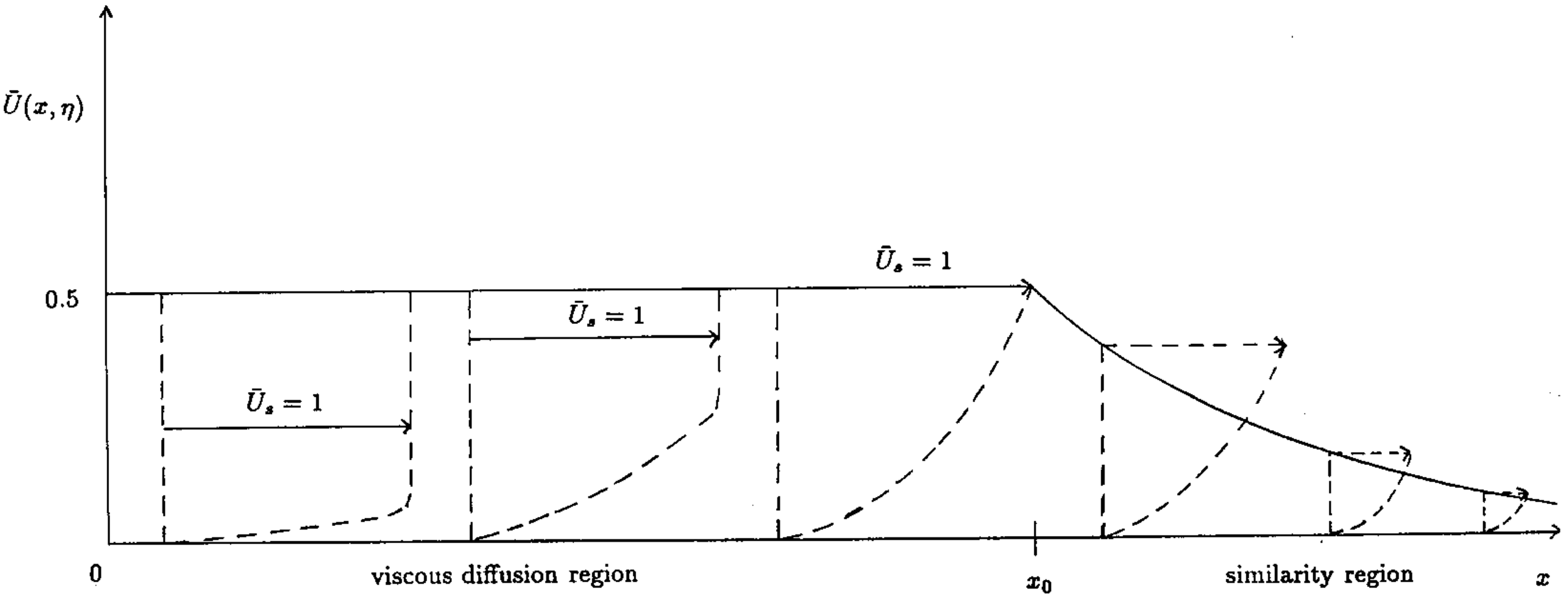

**Figure 4(a)**: Film thickness comparison of respective profiles

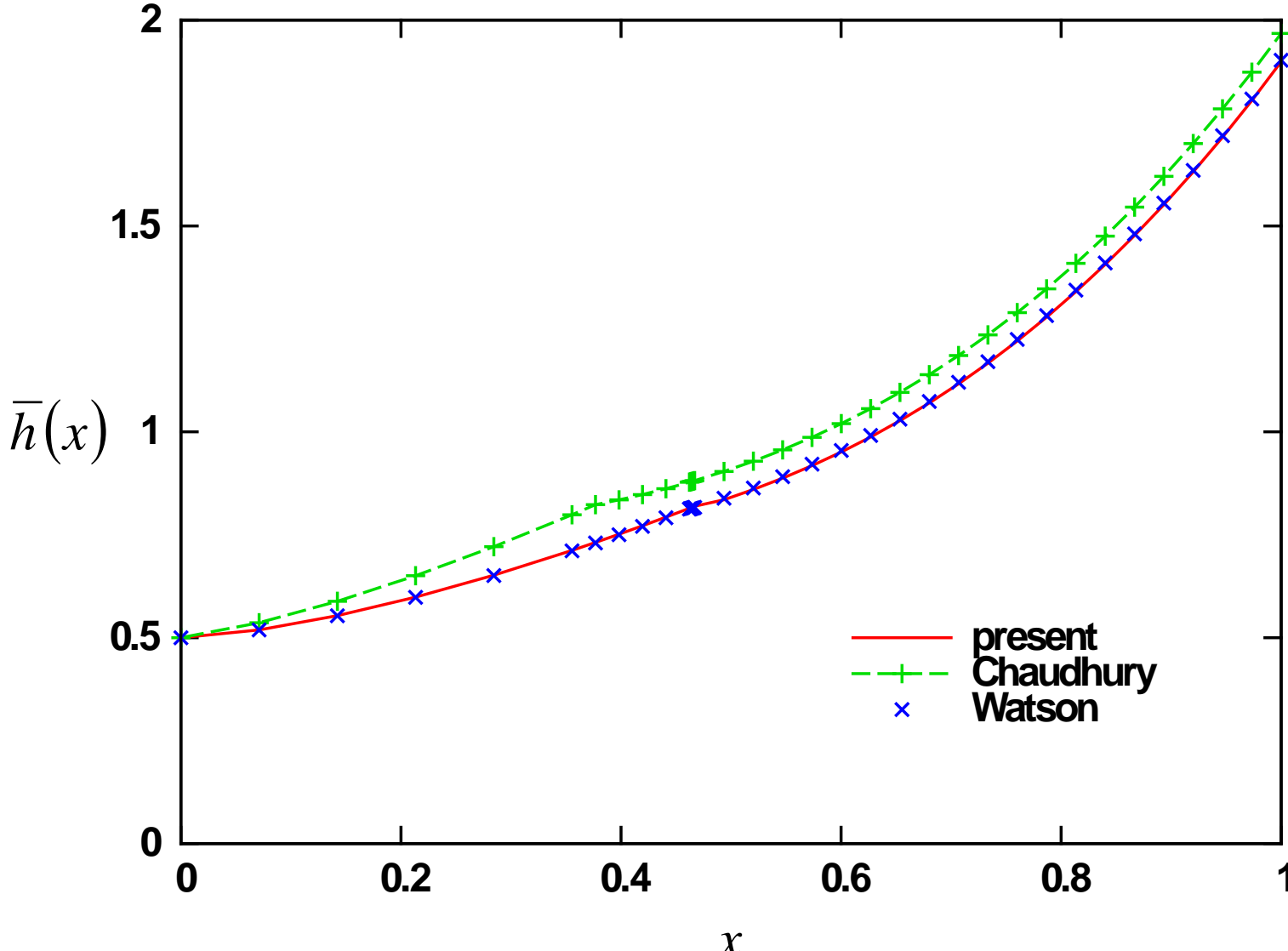

**Figure 4(b)**: Film thickness for numerical solution and present profile

2
1.5
$\bar{h}(x)$ 1
0.5
0
0 0.2 0.4 0.6 0.8 1
$x$

present profile
numerical solution

**Figure 5**: Thin film, viscous and thermal boundary layer thicknesses for $P_r$ = (a) $2$ (b) $5$ and (c) $10$ using present profile

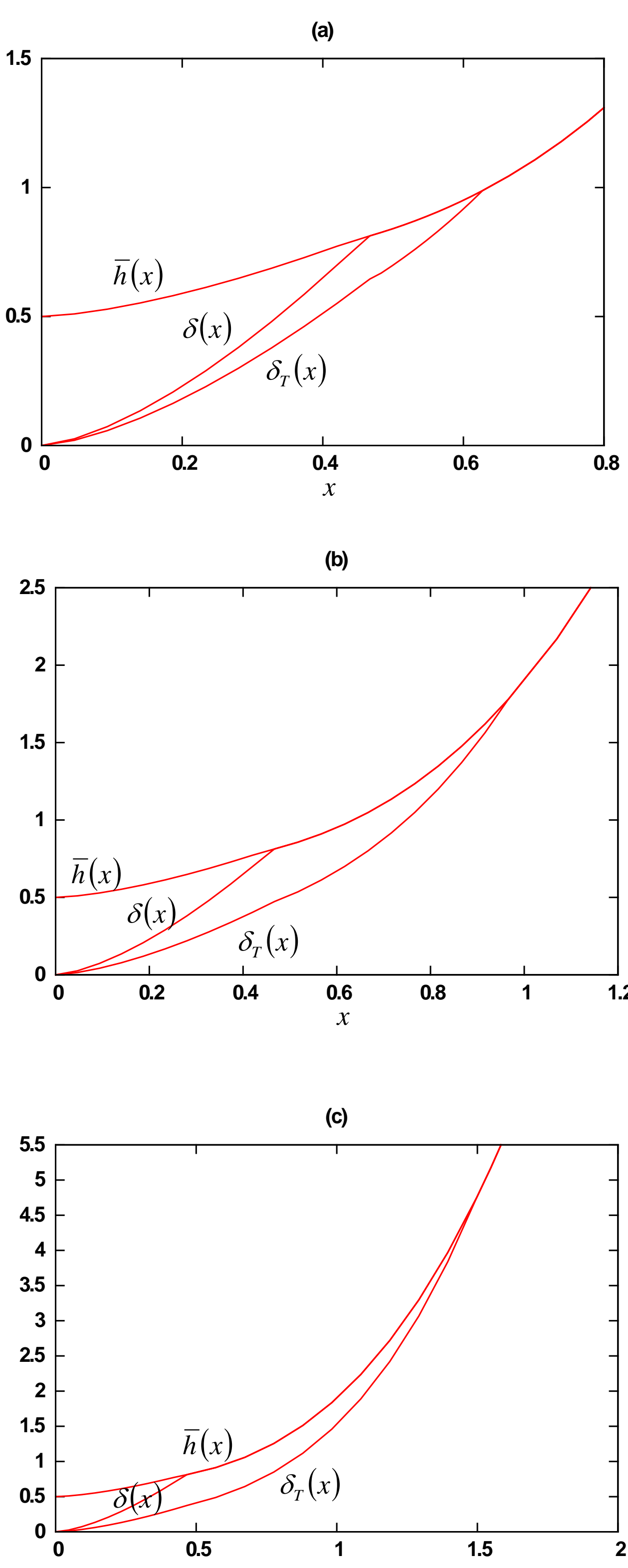

**Figure 6**: Temperature profile development within a deflection film

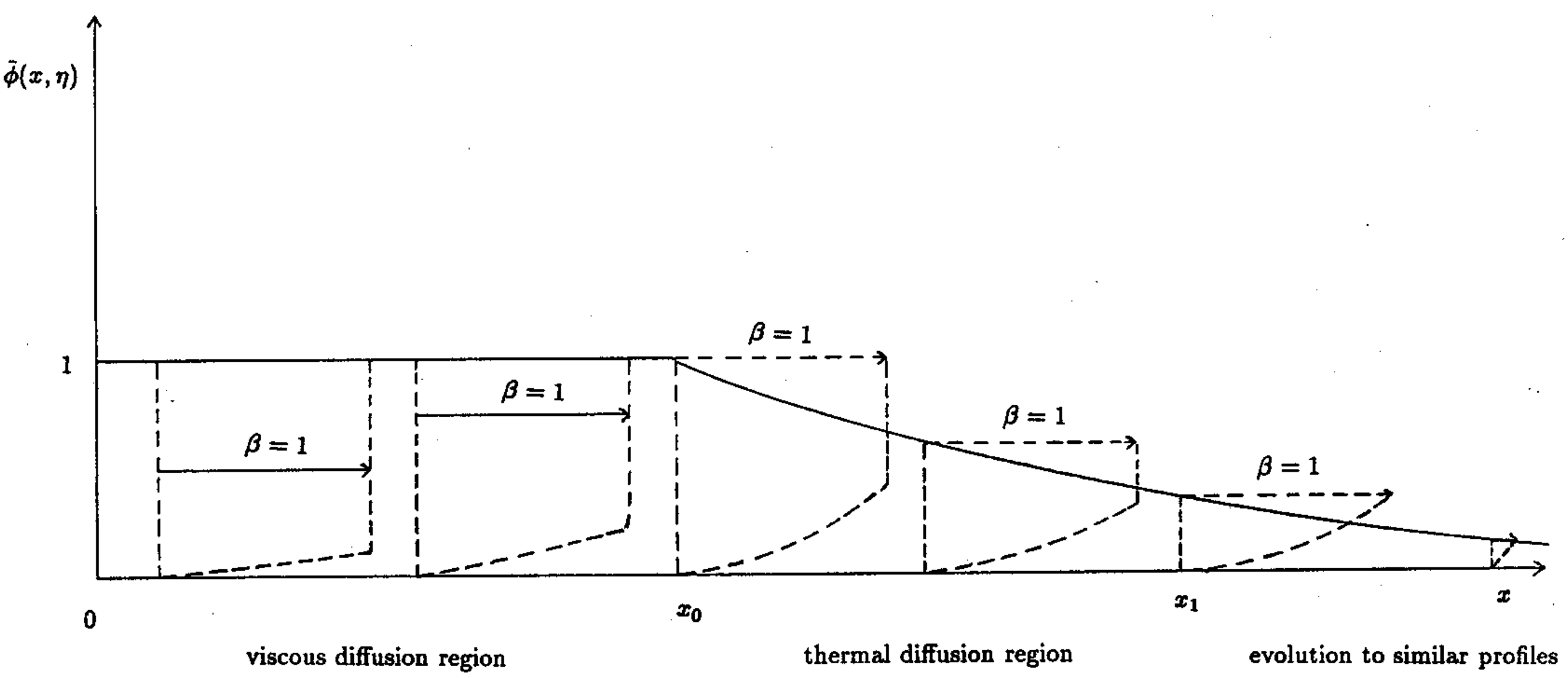

## 6.2 Skin friction and heat transfer coefficients

The elements of interest in engineering practice are the shear stress at the solid boundary *i.e.*, the skin friction and the rate of heat transfer at the boundary. The skin friction is defined as

$$\tau = \mu\left(\frac{\partial v_r}{\partial z}\right)_{z=0} \tag{67}$$

leading to the non-dimensional skin friction coefficient

$$\bar{\tau} = \left(\frac{H_0^2\mu^2}{\rho U_0^4}\right)^{\frac{1}{3}}\frac{\tau}{x} = \frac{\tau R_e^{\frac{2}{3}}}{\rho U_0^2 x} = \left(\frac{\partial \bar{U}}{\partial \bar{Y}}\right)_{\bar{Y}=0} . \tag{68}$$

From the approximations it gives

$$\bar{\tau}_p = \sqrt{\frac{37}{105x^3}} \qquad \text{in Region 1}$$

$$= \frac{81\sqrt{3}c^2}{8\pi^3\left(x^3 + l_p^3\right)^2} \qquad \text{in Regions 2 and 3} \tag{69}$$

$$\bar{\tau}_w = \sqrt{\frac{\left(72 + 39c - 19c^2\right)c}{420x^3}} \qquad \text{in Region 1}$$

$$= \frac{81\sqrt{3}c^3}{16\pi^3\left(x^3 + l_w^3\right)^2} \qquad \text{in Regions 2 and 3.} \tag{70}$$

The graphs of $\overline{\tau}(x)$ are plotted in Figure 7(a). The integrable square root singularity is consistent with the Blasius boundary layer equivalent.

**Figure 7(a)**: Skin friction of present and Chaudhury profiles

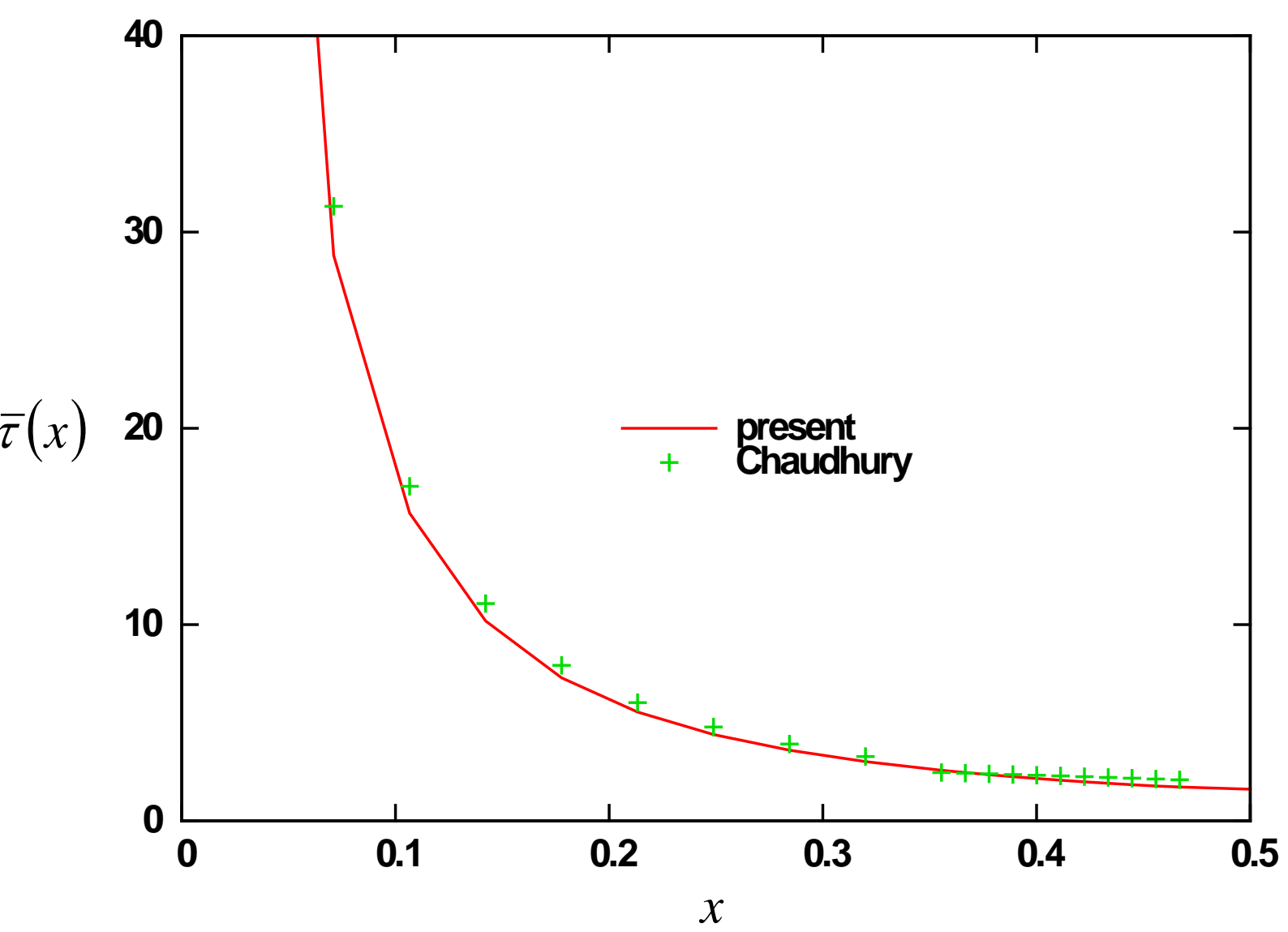

**Figure 7(b)**: Skin friction for numerical solution and present profile

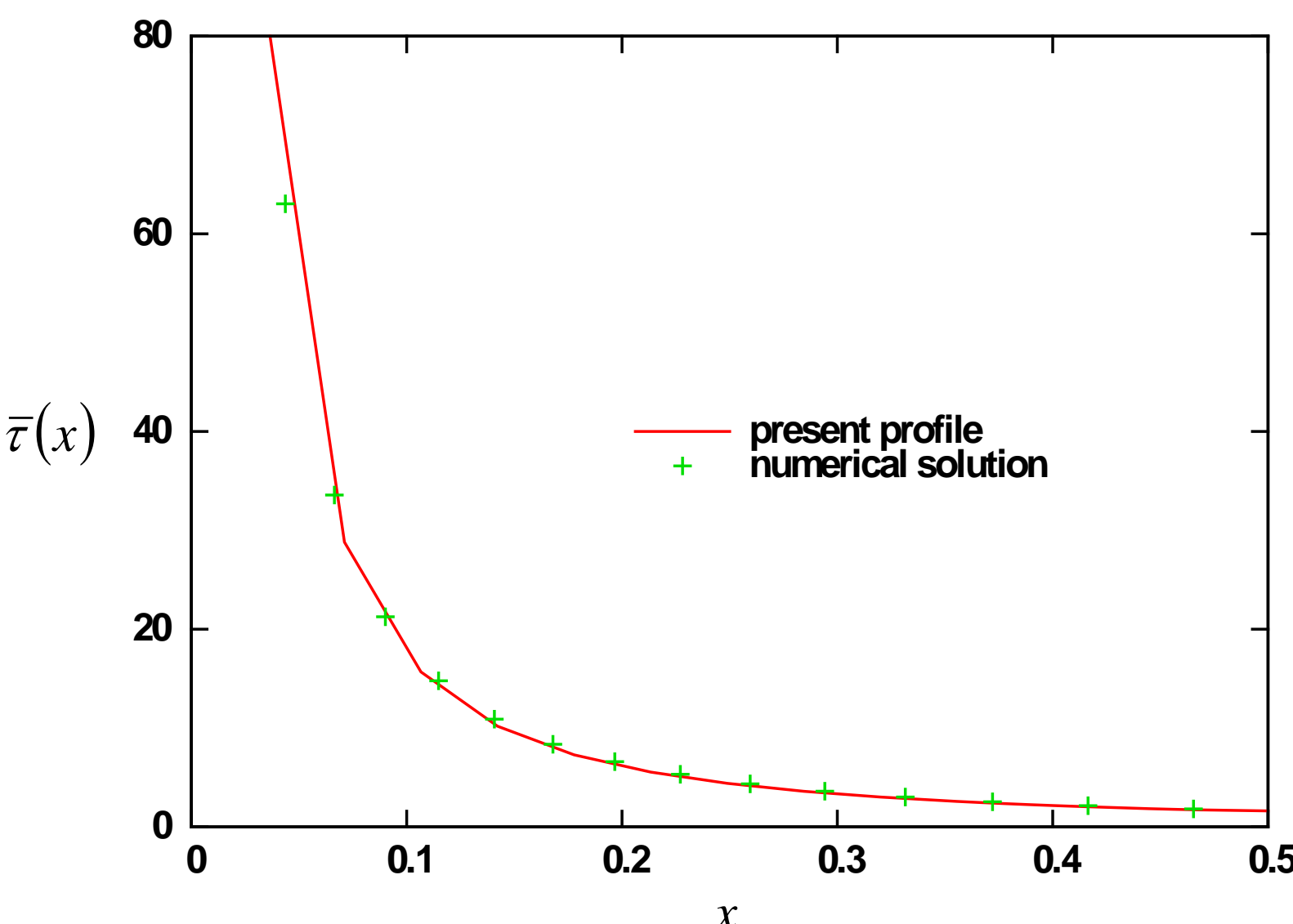

The most significant film cooling design factor is the heat transfer across the film. The heat transfer at the solid boundary is given by

$$q = -\kappa\left(\frac{\partial T}{\partial z}\right)_{z=0} = \frac{\kappa \Delta T R_e^{\frac{1}{3}} x}{H_0}\left(\frac{\partial \overline{\phi}}{\partial \overline{Y}}\right)_{\overline{Y}=0} \tag{71}$$

where $\Delta T = T_w - T_0$. The non-dimensional version of this is known as the Nusselt number defined as

$$N_u = \frac{qH_0}{\kappa \Delta T R_e^{\frac{1}{3}} x} = \left(\frac{\partial \overline{\phi}}{\partial \overline{Y}}\right)_{\overline{Y}=0}. \tag{72}$$

The results are

$$N_{u\,p} = \frac{1}{\Delta_p(P_r)}\sqrt{\frac{37}{105x^3}} \quad \text{in Region 1}$$

$$= \frac{1}{\Delta_p(x;P_r)}\frac{3\sqrt{3}}{\pi\left(x^3+l_p^3\right)} \quad \text{in Region 2} \tag{73}$$

$$= \frac{3\sqrt{3}}{\pi\left(x^3+l_p^3\right)}\left(\frac{x_{1\,p}^3+l_p^3}{x^3+l_p^3}\right)^{\frac{840}{367c^2P_r}} \quad \text{in Region 3}$$

and

$$N_{u\,w} = \frac{1}{\Delta_w(P_r)}\sqrt{\frac{\left(72+39c-19c^2\right)c}{420x^3}} \quad \text{in Region 1}$$

$$= \frac{1}{\Delta_w(x;P_r)}\frac{3\sqrt{3}c}{2\pi\left(x^3+l_w^3\right)} \quad \text{in Region 2} \tag{74}$$

$$= \frac{3\sqrt{3}c}{2\pi\left(x^3+l_w^3\right)}\left(\frac{x_{1\,w}^3+l_w^3}{x^3+l_w^3}\right)^{\frac{840}{c\left(360+111c+38c^2\right)P_r}} \quad \text{in Region 3.}$$

The predictions of $N_{u\,w}$ for a range of the Prandtl numbers are presented in Figure 8(a). The values of $\Delta_p(P_r)$ and $\Delta_w(P_r)$ have been obtained from the equations (56) and (57), respectively. $\Delta_w(x;P_r)$ satisfies the equation (61). Accordingly, $\Delta_p(x;P_r)$ satisfies the following equation

$$\Delta^3\left(3360-648\Delta^2+175\Delta^3\right) = \frac{50400}{c^2P_r}\ln\frac{x^3+l_p^3}{x_{1\,p}^3+l_p^3}+2887\,. \tag{75}$$

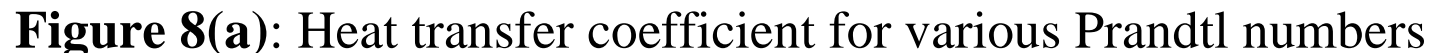

**Figure 8(a)**: Heat transfer coefficient for various Prandtl numbers

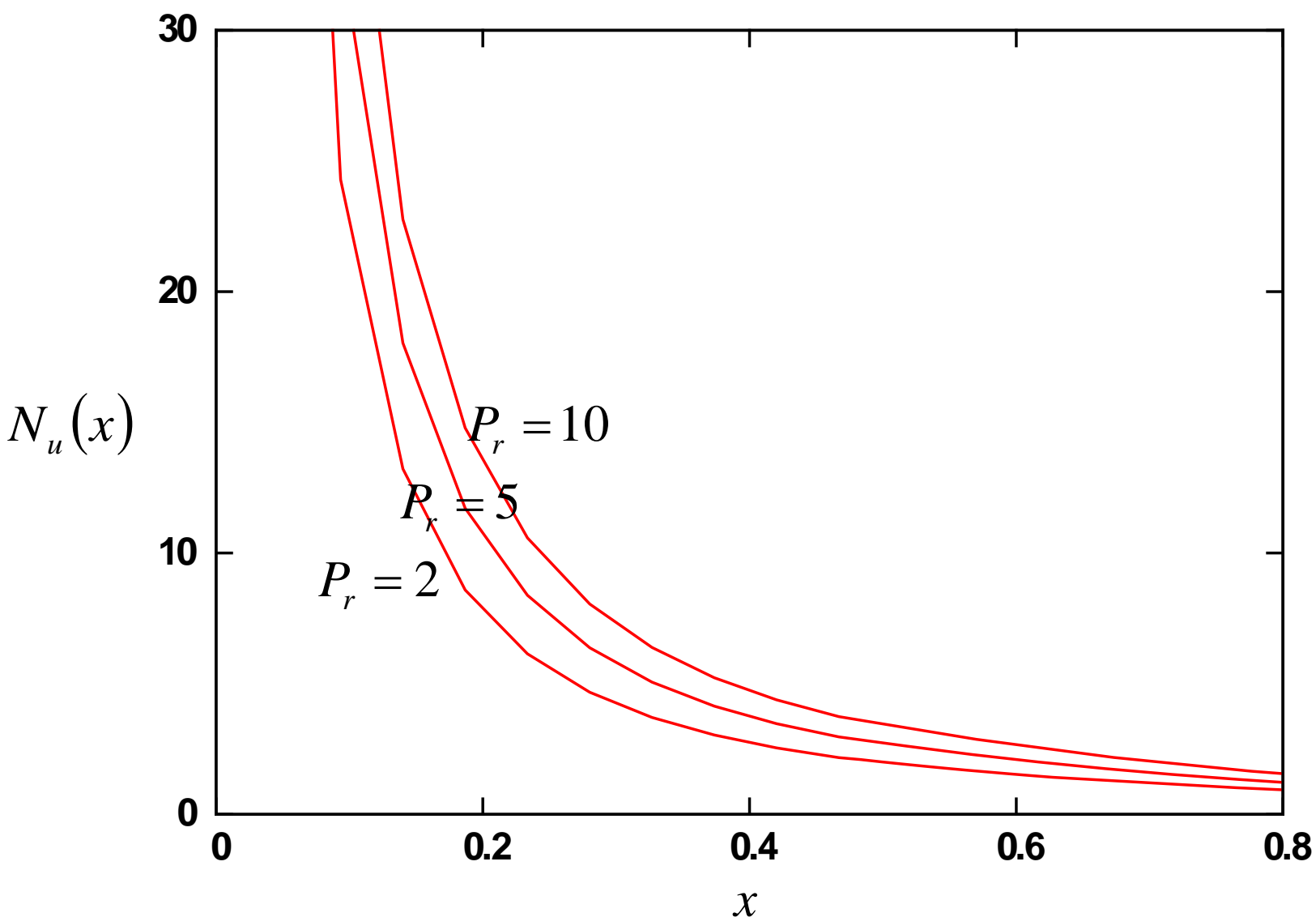

**Figure 8(b)**: Heat transfer coefficient for numerical solution and present profile at $P_r = 2$

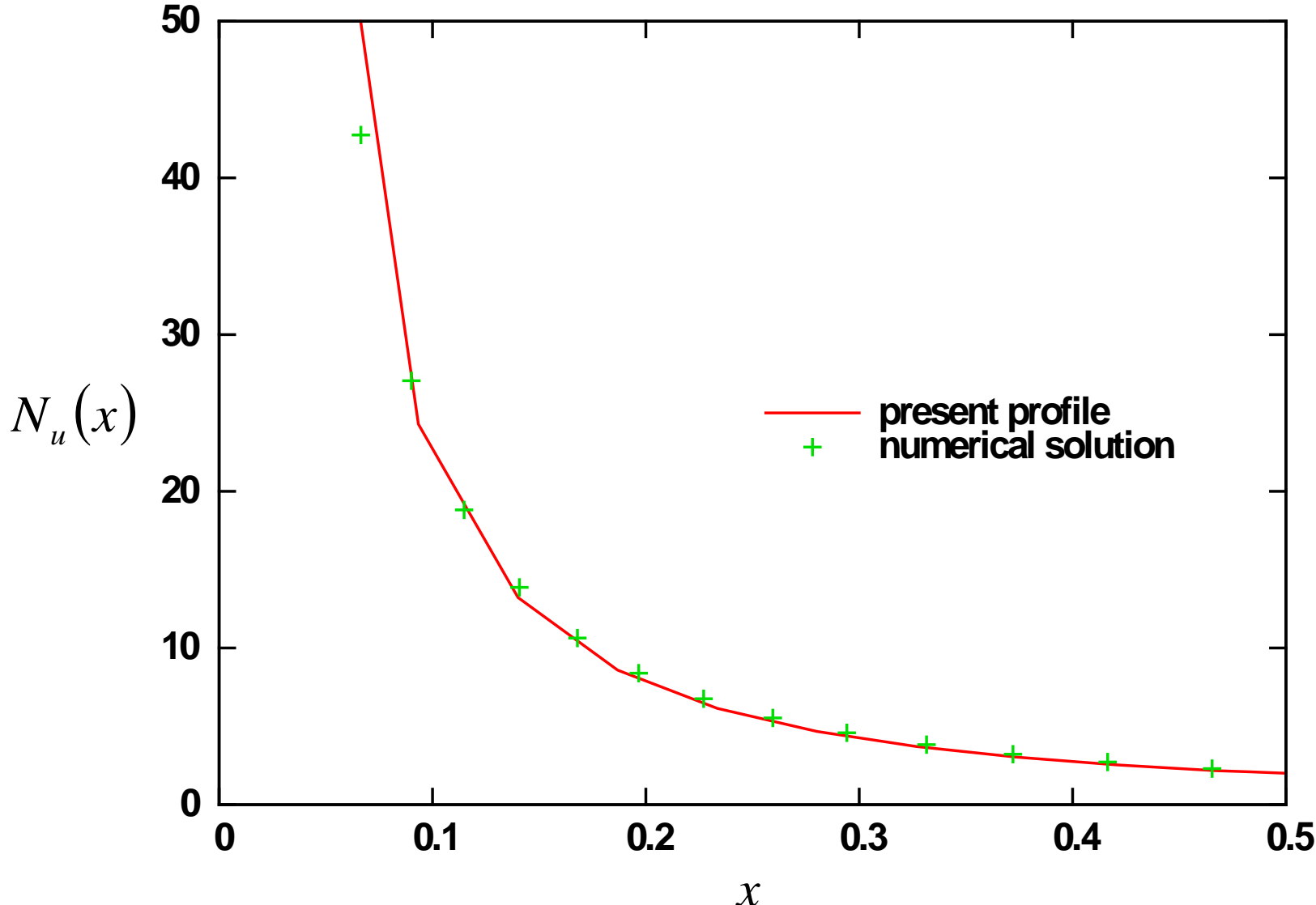

## 7 Numerical Solutions

The continuity equation (10) is eliminated by introducing a stream function $\psi$ defined by

$$\overline{U} = \frac{1}{x}\frac{\partial \psi}{\partial \overline{Y}}, \; \overline{V} = -\frac{1}{x}\frac{\partial \psi}{\partial x}. \tag{76}$$

Owing to the geometry, $\overline{H}(x)$ is singular at $x = 0$. To remove this singularity, $y$ and $\overline{h}(x)$ are introduced and given by

$$y = x\overline{Y},\ \overline{h}(x) = x\overline{H}(x). \tag{77}$$

Substituting equations (76) and (77) into (10)-(16) gives

$$\frac{\partial^3 \psi}{\partial y^3} = \left(\frac{1}{x^2}\right)\left(\frac{\partial \psi}{\partial y}\frac{\partial^2 \psi}{\partial x \partial y} - \frac{\partial \psi}{\partial x}\frac{\partial^2 \psi}{\partial y^2}\right) \tag{78}$$

$$\frac{\partial^2 \overline{\phi}}{\partial y^2} = \left(\frac{P_r}{x^2}\right)\left(\frac{\partial \psi}{\partial y}\frac{\partial \overline{\phi}}{\partial x} - \frac{\partial \psi}{\partial x}\frac{\partial \overline{\phi}}{\partial y}\right) \tag{79}$$

subject to boundary conditions

$$\psi = 0,\ \frac{\partial \psi}{\partial y} = 0,\ \overline{\phi} = 0 \quad \text{on} \quad y = 0,\ x \geq 0 \tag{80}$$

$$\psi = \frac{1}{2},\ \frac{\partial^2 \psi}{\partial y^2} = 0,\ \frac{\partial \overline{\phi}}{\partial y} = 0 \quad \text{at} \quad y = \overline{h}(x),\ x \geq 0 \tag{81}$$

$$\overline{h} = \frac{1}{2},\ \psi = y,\ \overline{\phi} = 1 \quad \text{on} \quad x = 0,\ 0 < y \leq \frac{1}{2} \tag{82}$$

where the initial condition (82) appears due to the original initial condition

$$H = \frac{H_0^2}{2r},\ v_r = U_0,\ T = T_0 \quad \text{on} \quad r = 0,\ 0 < z \leq \frac{H_0^2}{2r}. \tag{83}$$

In anticipation of the use of the Keller box method and its attractive extrapolation features, the differential system (78)-(82) is reformulated as the following first-order system

$$\frac{\partial \psi}{\partial y} = \overline{u}$$

$$\frac{\partial \overline{u}}{\partial y} = \overline{v}$$

$$\frac{\partial \overline{v}}{\partial y} = \left(\frac{1}{x^2}\right)\left(\overline{u}\frac{\partial \overline{u}}{\partial x} - \overline{v}\frac{\partial \psi}{\partial x}\right) \tag{84}$$

$$\frac{\partial \overline{\phi}}{\partial y} = \overline{w}$$

$$\frac{\partial \overline{w}}{\partial y} = \left(\frac{P_r}{x^2}\right)\left(\overline{u}\,\frac{\partial \overline{\phi}}{\partial x} - \overline{w}\,\frac{\partial \psi}{\partial x}\right)$$

whose boundary conditions are

$$\psi = 0,\ \overline{u} = 0,\ \overline{\phi} = 0 \quad \text{on} \quad y = 0,\ x \geq 0$$

$$\psi = \frac{1}{2},\ \overline{v} = 0,\ \overline{w} = 0 \quad \text{on} \quad y = \overline{h}(x),\ x \geq 0 \tag{85}$$

$$\overline{h} = \frac{1}{2},\ \psi = y,\ \overline{\phi} = 1 \quad \text{on} \quad x = 0,\ 0 < y \leq \frac{1}{2}.$$

The following coordinate transformation, what simultaneously maps the film thickness onto the unit interval and removes the Blasius singularity at the origin, is introduced

$$x = \xi^{\frac{2}{3}},\ \ y = \frac{\xi \eta \overline{h}}{\xi + 1 - \eta}\ .$$

The dependent variables are transformed as

$$\psi = \frac{\xi}{\xi + 1 - \eta} f\,,\ \overline{u} = \frac{u}{(1+\xi)^2}\,,\ \overline{v} = \frac{\xi + 1 - \eta}{\xi (1+\xi)^4} v,\ \overline{\phi} = \phi,\ \overline{w} = \frac{\xi + 1 - \eta}{\xi (1+\xi)^2} w,\ \overline{h} = (1+\xi)^2 h\,.$$

The equations to be solved now read

$$f_\eta = \frac{(1+\xi) h u}{\xi + 1 - \eta} - \frac{f}{\xi + 1 - \eta}$$

$$u_\eta = \frac{(1+\xi) h v}{\xi + 1 - \eta}$$

$$v_\eta = \frac{v}{\xi + 1 - \eta} - \frac{3\xi (1+\xi)^2 h u^2}{(\xi + 1 - \eta)^3} - \frac{3(1-\eta)(1+\xi)^3 h f v}{2(\xi + 1 - \eta)^4} + \frac{3\xi (1+\xi)^3 h}{2(\xi + 1 - \eta)^3}\left(u u_\xi - v f_\xi\right) \tag{86}$$

$$\phi_\eta = \frac{(1+\xi) h w}{\xi + 1 - \eta}$$

$$w_\eta = \frac{w}{\xi + 1 - \eta} - \frac{3P_r(1-\eta)(1+\xi)^3 hfw}{2(\xi + 1 - \eta)^4} + \frac{3P_r\xi(1+\xi)^3 h}{2(\xi + 1 - \eta)^3}\left(u\phi_\xi - wf_\xi\right)$$

subject to

$$f = 0,\ u = 0,\ \phi = 0 \quad \text{on} \quad \eta = 0,\ \xi \geq 0$$

$$f = \frac{1}{2},\ v = 0,\ w = 0 \quad \text{on} \quad \eta = 1,\ \xi \geq 0 \tag{87}$$

$$h = \frac{1}{2},\ f = f_0(\eta),\ \phi = \phi_0(\eta) \quad \text{on} \quad \xi = 0,\ 0 < \eta \leq 1$$

where the initial profiles $f_0(\eta)$ and $\phi_0(\eta)$ are found by putting $\xi = 0$ and $h = \frac{1}{2}$ into (86) and solving, subject to conditions $f = u = \phi = 0$ at $\eta = 0$ and $u = 1$, $\phi = 1$ at $\eta = 1$. The parabolic system of equations and boundary conditions (86)-(87) has been solved by marching in the $\xi$-direction using a modification of the Keller box method. A non-uniform grid is placed on the domain $\xi \geq 0$, $0 \leq \eta \leq 1$ and the resultant difference equations are solved by Newton iteration. Solutions are obtained on grids of varied sizes, and the Richardson's extrapolation is used to produce results of high accuracy. A full account of the numerical method and the details of implementation are beyond the scope of this paper and will be reported separately [13]. The detailed numerical method procedure for this case is fully discussed in [1]. For the axisymmetric flat plate in this paper, the relevant physical parameters should be chosen as

$$F(x) = 0,\ G(x) = \frac{1}{x},\ x_s = +\infty,\ \gamma = \frac{1}{2},\ \alpha = \frac{2}{3},\ \beta = 2.$$

The solution scheme was successfully evaluated against previously reported results [14-19].

# 8 Results

A typical run has a coarse grid of dimension $60 \times 48$ in the ($\xi$, $\eta$) domain with each cell being divided into $1, 2, 3$ and $4$ sub-cells, respectively. Because of the coordinate singularity at $\xi = 0$, $\eta = 1$, a non-uniform grid is employed and given by $\xi = \frac{1}{3}\sinh\left[\overline{\xi}^{1.5}\left(1 + \overline{\xi}^{1.5}\right)\right]$, $\eta = 1 - (1 - \overline{\eta})^{1.5}$ where $\overline{\xi}$ and $\overline{\eta}$ are uniform. When $\Delta\overline{\xi} \equiv 0.044618955$ and $\Delta\overline{\eta} \equiv \dfrac{1}{47}$, this gives $\Delta\xi \sim 0.004$ and $\Delta\eta \sim 0.003$ near the singularity, which is sufficiently small to give good accuracy, and this enabled us to integrate as far as $\xi \sim 10^9$, which is necessary for the profile at infinity to be determined with sufficient accuracy. From the convergence of the extrapolation process the absolute error is $9\times 10^{-7}$. A typical set of numerical data is presented in Table 3.

**Table 3**: Film thickness, free surface velocity, and temperature for an axisymmetric flat plate with $P_r = 2$

| $x$ | film thickness $\overline{h}(x)$ | free surface velocity $\overline{u}(x,\overline{h}(x))$ | free surface temperature $\overline{\phi}(x,\overline{h}(x))$ |
|---|---|---|---|
| 0.000 | 0.500 | 1.000 | 1.000 |
| 0.115 | 0.539 | 1.000 | 1.000 |
| 0.197 | 0.587 | 1.000 | 1.000 |
| 0.294 | 0.659 | 1.000 | 1.000 |
| 0.416 | 0.767 | 0.981 | 0.999 |
| 0.520 | 0.875 | 0.911 | 0.989 |
| 0.733 | 1.191 | 0.682 | 0.891 |
| 1.072 | 2.206 | 0.368 | 0.660 |
| 1.669 | 6.338 | 0.128 | 0.389 |
| 1.968 | 9.930 | $8.188\times 10^{-2}$ | 0.310 |
| 2.817 | 27.742 | $2.931\times 10^{-2}$ | 0.185 |
| 5.228 | $1.735\times 10^{2}$ | $4.685\times 10^{-3}$ | $7.372\times 10^{-2}$ |
| 10.791 | $1.520\times 10^{3}$ | $5.348\times 10^{-4}$ | $2.484\times 10^{-2}$ |
| 25.010 | $1.892\times 10^{4}$ | $4.297\times 10^{-5}$ | $7.031\times 10^{-3}$ |
| 46.931 | $1.250\times 10^{5}$ | $6.504\times 10^{-6}$ | $2.734\times 10^{-3}$ |
| $1.347\times 10^{2}$ | $2.957\times 10^{6}$ | $2.749\times 10^{-7}$ | $5.620\times 10^{-4}$ |
| $1.073\times 10^{3}$ | $1.495\times 10^{9}$ | $5.438\times 10^{-10}$ | $2.500\times 10^{-5}$ |
| $1.321\times 10^{4}$ | $2.784\times 10^{12}$ | $2.920\times 10^{-13}$ | $6.00\times 10^{-7}$ |
| $1.385\times 10^{5}$ | $3.212\times 10^{15}$ | $2.531\times 10^{-16}$ | 0.000 |
| $1.000\times 10^{6}$ | $1.209\times 10^{18}$ | $6.723\times 10^{-19}$ | 0.000 |

Figures 2(b), 4(b), 7(b), 8(b) and 9(a) show excellent agreement between the full numerical solutions and the theoretical results obtained using an assumed present quartic velocity profile (50). Figure 9(b) shows free surface temperature for various Prandtl numbers. As $P_r$ increases, the temperature decrease becomes more gradual.

**Figure 9(a)**: Free surface temperature for numerical solution and present profile at $P_r = 2$

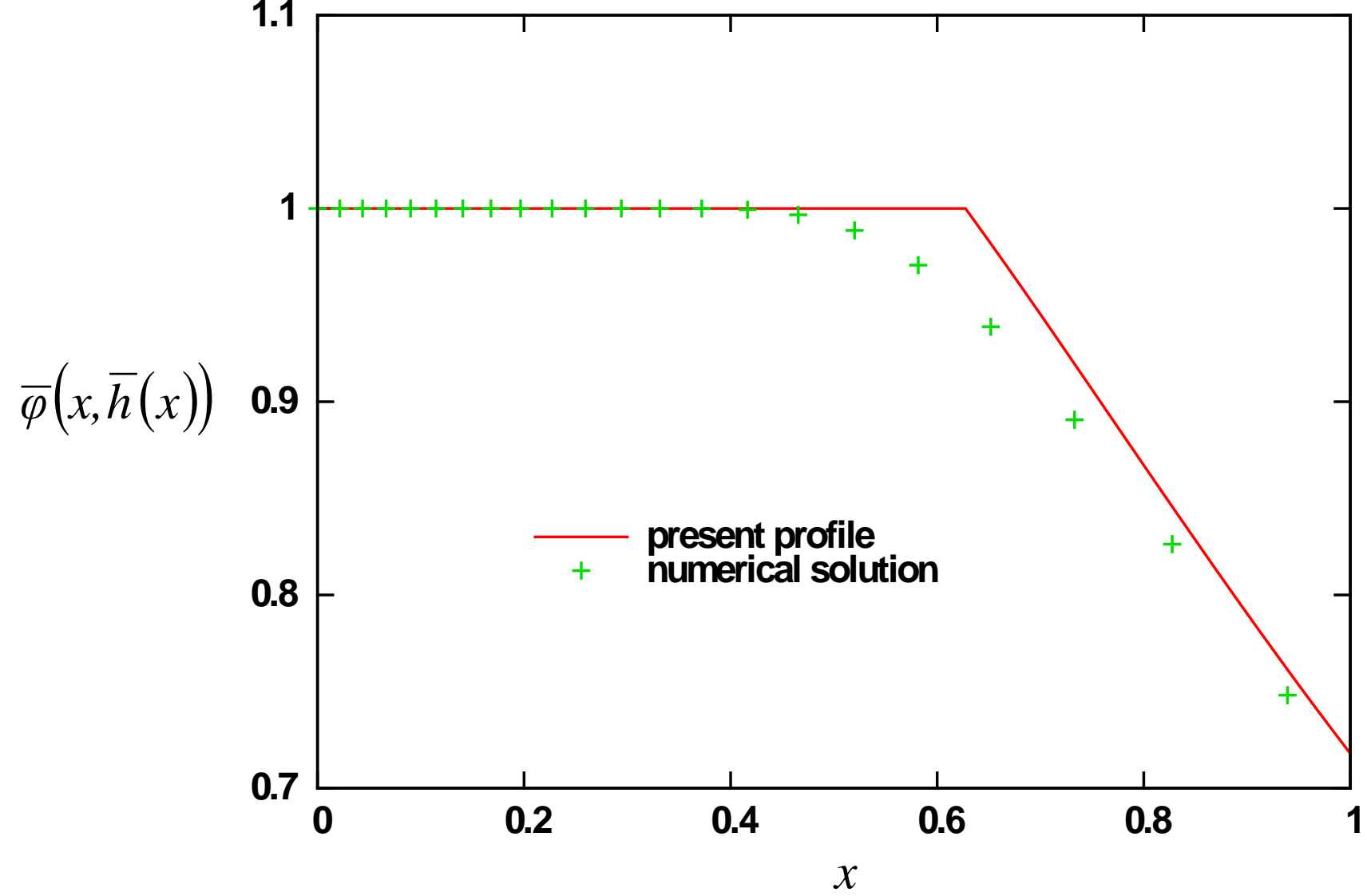

**Figure 9(b)**: Free surface temperature for various Prandtl numbers

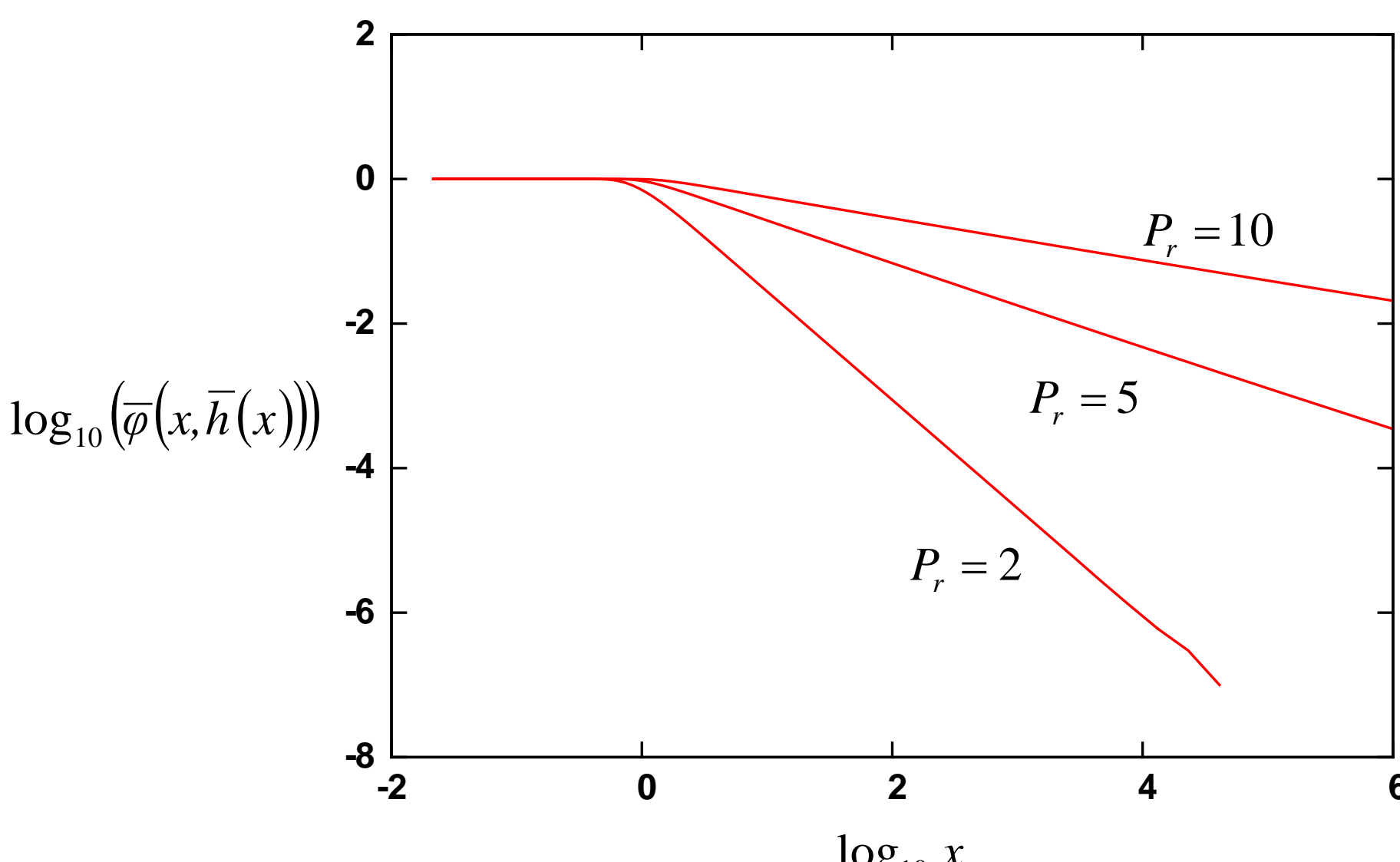

## 9 Concluding Remarks

Both numerical and approximate solutions for the flow of a cold axisymmetric vertical jet against a horizontal, flat plate have been obtained. As the prediction of skin friction and heat transfer characteristics is concerned, the level of agreement in estimating film thickness, velocity and temperature distributions is remarkably good insofar. Although it is valuable to have demonstrated a successful, robust Keller-box algorithm, it is also noteworthy that the approximation may provide a satisfactory methodology for the assessment of practical configurations, sufficient for the purposes of engineering practice.